\documentclass{ws-procs975x65}

\begin{document}

\title{NON-EXISTENCE OF STATIONARY TWO-BLACK-HOLE CONFIGURATIONS}

\author{J\"org Hennig}

\address{Max-Planck-Institut f\"ur Gravitationsphysik,
Albert-Einstein-Institut,\\
Am M\"uhlenberg 1, D-14476 Potsdam, Germany\\
E-mail: pjh@aei.mpg.de}

\author{Gernot Neugebauer}

\address{Theoretisch-Physikalisches Institut,
Friedrich-Schiller-Universit\"at Jena,\\
Max-Wien-Platz 1, D-07743 Jena, Germany\\
E-mail: G.Neugebauer@tpi.uni-jena.de}

\begin{abstract}
We resume former discussions of the question, whether the spin-spin 
repulsion and the gravitational attraction of two aligned sub-extremal
black holes 
can balance each other. To answer the question we formulate a boundary 
value problem for two separate \mbox{(Killing-)}
horizons and apply the inverse 
(scattering) method to solve it. Making use of a universal inequality
between angular momentum and horizon area that has to be satisfied
by every sub-extremal black hole, we prove the
non-existence of the equilibrium situation in question.
\end{abstract}


\bodymatter

\section{Introduction}

This talk is meant to contribute to the present discussion about the
existence or non-existence of stationary equilibrium configurations
consisting of separate bodies at rest. In Newtonian theory it is
a classical result that there exists
no static $n$-body configuration (with bodies
separated by a plane and with $n>1$).
Recently, a similar statement was
shown in the context of General relativity: Beig and Schoen \cite{Beig}
were able to
prove a non-existence theorem for a reflectionally symmetric
\emph{static} $n$-body configuration.
In the meantime, Beig \emph{et al.}\cite{Beig2}
have been able to extend this
non-existence theorem to symmetric configurations with
\emph{anti-aligned} spins. 

However, as indicated by post-Newtonian approximations, anti-aligned
spins enhance the omnipresent mass attraction. Hence it could be desirable
to study examples of \emph{aligned} spins that could generate repulsive
effects.

As a characteristic example for such configurations we investigate the
possibility of equilibrium between two aligned rotating axisymmetric
\emph{black holes}.
We will present and review a chain of old and new arguments which
finally forbid this equilibrium situation.
The following considerations are based on
another article \cite{Neugebauer2009}.

\section{The double-Kerr-NUT solution}

Interestingly, there exists an exact solution to the Einstein
equations that was expected to solve the two-black-hole equilibrium
problem \cite{Dietz,Hoenselaers1983,Hoenselaers1984,Kihara1982,Kramer1980,Kramer1986,Krenzer,Manko2000,Manko2001}
--- the \emph{double-Kerr-NUT solution}, first investigated
by Kramer and Neugebauer \cite{Kramer1980}. This solution is a particular case
of a more general solution which was constructed
by applying an $N$-fold \emph{B\"acklund transformation}\footnote{The
B\"acklund transformation is a particular method from soliton theory
that creates new solutions to nonlinear equations from a previously
known one.}
to an arbitrary seed solution \cite{Neugebauer1980a}.
The double-Kerr-NUT solution can be obtained as the special case of a
two-fold ($N=2$) B\"acklund transformation
applied to Minkowski spacetime \cite{Neugebauer1980b}. Since a single B\"acklund
transformation generates the Kerr-NUT solution that contains, by a
special choice of its parameters, the Kerr black hole solutions and
since B\"acklund transformations act as a ``nonlinear superposition
principle'', the double-Kerr-NUT solution was considered to be a good
candidate for the solution of the two-horizon problem.

However, there was no argument requiring
that this particular solution be the \emph{only} candidate.
The deciding step to remove this objection was to formulate and discuss
a boundary value problem for two separate horizons.
Using the \emph{inverse scattering method
of soliton theory}\footnote{Hereby, an
associated linear problem is analyzed, whose 
integrability conditions are equivalent to the non-linear field
equations.}
we were able to show that the solution of this boundary value problem
necessarily leads to (a subclass
of) the double-Kerr-NUT family of solutions.
As a consequence, we can indeed make use of former
discussions\cite{Manko2000,Manko2001} of the double-Kerr-NUT solution.
Details of the solution of
the boundary value problem can be found in Neugebauer
\emph{et al.}\cite{Neugebauer2003,Neugebauer2009}
These articles make use of a particular form of the linear
problem~\cite{Neugebauer1980b} (see Eq.~(1) of this reference)
which is equivalent to the linear problem of
Belinski and Zakharov\cite{Belinski}.

\section{A universal inequality for sub-extremal black holes}

In the following we investigate whether the double-Kerr-NUT solution 
can describe equilibrium between two \emph{sub-extremal} black holes
(where we define sub-extremality by existence of trapped surfaces in
every sufficiently small interior neighborhood of the event
horizon\cite{Booth}). The possibility of two-black-hole equilibrium
configurations involving \emph{degenerate} black holes will be studied in a
forthcoming article\cite{Hennig2010}.

With methods from the calculus of variations it can be
shown\cite{Hennig2008a} that every sub-extremal black hole satisfies
the inequality
\begin{equation}\label{ineq}
 8\pi|J|<A
 \end{equation}
between angular momentum $J$ and horizon area $A$.
With the explicit formulae for the quantities $A$ and $J$ for the two
gravitational objects (black hole candidates) described by the
double-Kerr-NUT solution we can test whether inequality \eqref{ineq} is
satisfied for both objects. However,
it turns out
that for all choices of the free parameters at least one of these two
objects violates the inequality.
Hence we conclude that \emph{an equilibrium
between two sub-extremal black holes is impossible}.

\section*{Acknowledgments}
This work was supported by the Deutsche
For\-schungsgemeinschaft (DFG) through the
Collaborative Research Centre SFB/TR7
``Gravitational wave astronomy''.


\end{document}